# Semi-strong Efficient Market of Bitcoin and Twitter: an Analysis of Semantic Vector Spaces of Extracted Keywords and Light Gradient Boosting Machine Models

Fang Wang[a,1,*], Marko Gacesa[b,2]

[a]*Rensselaer Polytechnic Institute, 110 Eighth Street, Troy, NY, United States*
[b]*Physics Department, Khalifa University, P.O. Box 127788, Abu Dhabi, United Arab Emirates*

**Abstract**

This study extends the examination of the Efficient-Market Hypothesis in Bitcoin market during a five-year fluctuation period, from September 1 2017 to September 1 2022, by analyzing 28,739,514 qualified tweets containing the targeted topic "Bitcoin". Unlike previous studies, we extracted fundamental keywords as an informative proxy for carrying out the study of the EMH in the Bitcoin market rather than focusing on sentiment analysis, information volume, or price data. We tested market efficiency in hourly, 4-hourly, and daily time periods to understand the speed and accuracy of market reactions towards the information within different thresholds. A sequence of machine learning methods and textual analyses were used, including measurements of distances of semantic vector spaces of information, keywords extraction and encoding model, and Light Gradient Boosting Machine (LGBM) classifiers. Our results suggest that 78.06% (83.08%), 84.63% (87.77%), and 94.03% (94.60%) of hourly, 4-hourly, and daily bullish (bearish) market movements can be attributed to public information within organic tweets.

*Keywords:* Efficient-Market Hypothesis; Twitter; Bitcoin; LightGBM; GloVe Semantic Vector Spaces

## 1. Introduction

The Efficient-Market Hypothesis (EMH) (Fama 1970) has been arguable within different markets. Since Bitcoin (Nakamoto 2008) has became one of the most volatile assets in financial markets, researchers have extended debates of market efficiency to cryptocurrency markets. Garcia et al. (2014) was first to reject the EMH in Bitcoin, followed by Urquhart (2016), Cheung et al. (2015), Cheah and Fry (2015), Bariviera (2017) and other researchers. These studies utilized various approaches, including intrinsic rate of return, log-periodic power law, Hurst exponent (Hurst 1951), and the nil-intrinsic valuation, among others, to further approve the argument. In contrast, Bartos et al. (2015), Nadarajah and Chu (2017), Tiwari et al. (2018), and Vidal-Tomás and Ibañez (2018), have demonstrated and argued in favor of either a weak form or a semi-strong form of the EMH throughout Bitcoin market reactions between 2013 and 2017. Majority of these studies have focused on the early development stage of Bitcoin, from 2009 to 2018, when increases were more favored in most market movements. Regardless of statistical methods and empirical data, the question of how information is disseminated among investors and at what exact speed during more the period of greater fluctuations, namely after 2018, still needs to be addressed.

Twitter has established itself as an efficient tool for examining when and what type of information are available on financial markets, since existing effects of networking and information exchanges could be beneficial among investors(Bartov et al. 2018). Significant relationships between Twitter and financial market have been discovered by Bollen et al. (2011), Piñeiro-Chousa et al. (2018), Sun et al. (2016) and other researchers. Moreover, impacts of volumes and sentiments of tweets on the next-day volatility of Bitcoin market have been studied (Shen et al. 2019; Kaminski 2014). Regardless of these studies, fundamental informative elements – words – comprising the tweets that could signal different market movements and be used to determine the extent of market efficiency in the Bitcoin market still remain unknown. The importance of similar standards of analyses being conducted by researchers with adequate theoretical justification should be noted. We believe that a suitable approach to examining the relationship between public information and Bitcoin price changes will be important for the study of market efficiency in this field.

In this paper, we address above questions and examine to what extent are the market movements impacted by preceding information revealed on Twitter. Under the presumption of the EMH, if Bitcoin market is efficient, market prices should reflect on the public discussions/information available on Twitter. We also explore how accurate reflections vary pursuant to different time thresholds and properties of information. For purposes of this study, we collected 28,739,514 qualified tweets related to the topic "Bitcoin" over the period with the most market fluctuations, from September 1 2017 to September 1 2022, and analyzed the information that they convey to identify the influences towards succeeding market movements. To realize this procedure, we set models on the basis of three groups of possible outcomes within different time intervals: hourly, 4-hourly, and daily sequential market movements. We then implemented a sequence of machine learning methods and textual analyses, including measurements of distances of semantic vector spaces, keywords extractions and encoding model, and Light Gradient Boosting Machine (LGBM) classifier, to search for the effects. Our predictive model suggests that above 95% of daily movements can be attributed to public information with organic contents.

The contribution of this work is as follows. First, we further extend the discussion of the EMH in Bitcoin market by examining the most fluctuating period and the speed of market reactions. We prove that the market is semi-strong efficient with respect to Twitter discussions. Here, we do not limit the discussion of the EMH to a daily time series approach but instead include high-frequency market actions data, namely hourly and 4-hourly, as suggested by Bariviera and Merediz-Solà (2021). Second, instead of conducting analyses on volumes or sentiments of information, we deploy semantic distances from vector spaces and feature vector encoding system to distinguish market effects in correspondence with different informative contents. The preceding information represented by certain keywords confirms an efficient market that is able to reflect general information in normal communication without involving special messages from specific trigger events (Vidal-Tomás and Ibañez 2018). Here, the most conclusive results are obtained by LGBM, which as an advanced machine learning approach would be used for similar future studies. Third, we demonstrate that market movements are more sensitive towards negative information, leading an initial exploration of unique properties related to market development of Bitcoin.

The paper is organized as follows. Following the introduction given in Section 1, Section 2 includes literature review on the EMH in Bitcoin and Twitter. In Section 3, we present the data and research methodology. In Section 4 we present and analyze the results. We discuss the implication and future work in Section 5.

## 2. Literature Review

Previous research explains the EMH as informational efficiency in a market that occurs to a degree if "prices reflect full information" (Fama 1970). The EMH is commonly understood to hold in three forms. Namely, weak-form efficiency stands for unpredictable return regardless of all and any information inhibited in past prices. Semi-strong efficiency represents the situation where prices reflect all information easily accessible to the public. Finally, strong-form efficiency stands for present prices reflecting all public and private (hidden) information.

Initial significant studies of market efficiency in the Bitcoin market were mostly conducted on analyses of price patterns. Urquhart (2016) collected daily price data from Aug 1 2010 to Jul 31 2016, examined whether prices followed a random walk, and concluded that Bitcoin was initially an inefficient market that might have been in the process of moving towards an efficient market after Aug 2013. In a similar perspective, Nadarajah and Chu (2017) employed the same data as Urquhart (2016) and used eight testing methods to demonstrate that a simple power transformation of the Bitcoin returns satisfied the EMH. Bariviera (2017) deployed Hurst exponent and used daily price data covering the period from Aug 18 2011 to Feb 15 2017, to find that daily return time series became more efficient across the time domain even though the volatility exhibited long memory during the entire time period. Tiwari et al. (2018) revisited previous research and collected Bitcoin market data from Jul 18 2010 to Jun 16 2017. Their observations signaled efficiency of the Bitcoin market, with exceptions of April-August in 2013 and August-November in 2016. Jiang et al. (2018) investigated time-varying long memory in the period from Dec 1 2010 to Nov 30 2017 using a new efficiency index and confirmed the presence of long memory in the Bitcoin market as well as declined the change of the market over time towards greater efficiency. Along the same line, Cheah et al. (2018) employed daily price data concerning the period Nov 27 2011 to Mar 17 2017 from 5 different regional markets (Europe, USA, Australia, Canada, and the United Kingdom) and have found that medium-to-high inefficiencies are present across Bitcoin regional markets and its long-memory characteristics permit trading profits. Consequently, previous researchers were not able to reach a consensus about the EMH in Bitcoin market since their analyses were based solely on price data taken during the initial development stage of Bitcoin. Moreover, inconsistencies in the time series of prices of cryptocurrency market have been already addressed as an issue (Vidal-Tomás 2021; Alexander and Dakos 2020), leading to necessities for more comprehensive data and methods used in relevant research.

The intensely bullish market of 2017 brought about increasing academic attention in investigating the impacts on the market caused by different sources of information. For example, Al-Yahyaee et al. (2018) set the research period from Jul 18 2010 to Oct 31 2017 and simultaneously recorded economic signals from gold, stock, and foreign exchange markets in order to assess the efficiency of Bitcoin market in comparison to other more established markets. They found stronger long-memory features and multifractality of Bitcoin market, which drive the market to be more inefficient than the other analyzed markets. Similarly, Kinateder and Papavassiliou (2021) could not affirm the EMH in Bitcoin market due to the discovery of a relationship between the calendar effects (*i.e.*, holidays) and price movements. Presumptions of these studies to reject the EMH were grounded on somewhat irrelevant signals or effects, instead of impacts from public information closely related to Bitcoin. In contrast, the debate is in favor of confirming at least some level of the EMH when based solely on meaningful information. Vidal-Tomás and Ibañez (2018) applied market events as informative leverage and argued that based on the data from Sept 13 2011 to Dec 17 2017, Bitcoin became more efficient over time in relation to its own market events. Wang et al. (2022) introduced an index of cryptocurrency environmental attention (ICEA) to capture the relative extent of media discussions surrounding the environmental impact of cryptocurrencies and found out that Bitcoin has the strongest reactions from the ICEA variation shocks. Hossain (2021) also suggested that the information asymmetry in cryptocurrencies market could play a role in pricing movement. Researchers, such as Kim et al. (2016); Karalevicius et al. (2018); Kraaijeveld and De Smedt (2020), conducted studies on sentiment analysis of relevant information on social media, demonstrating how information drove volatility of Bitcoin market. It is worth mentioning that the only exception happened in 2014 (Garcia et al. 2014), the early phase of Bitcoin market, due to the discovery of two positive feedback loops in the market that originated price bubbles in absence of exogenous stimuli. Furthermore, Lucey et al. (2022) replied on news coverage and developed a new Cryptocurrency Uncertainty Index (UCRY), capturing uncertainty beyond Bitcoin. Hence, market behaviors corresponding to information are not consistent over time and the uncertainty of the EMH in Bitcoin market after 2017 still remains an open question.

In our view, a crucial element to validate the EMH is the manner in which information is publicly distributed. Blankespoor et al. (2014); Bartov et al. (2018) and other researchers have confirmed that Twitter could serve as a medium of information dissemination and have effects on market efficiency in financial markets. Numerous studies have focused on sentiment analysis of tweets for purposes of illustrating the EMH in stock market (Pagolu et al. 2016; Piñeiro-Chousa et al. 2018; Cao et al. 2022), because "the public mood and sentiment can drive stock market values as much as news" (Bollen et al. 2011). This approach has been adopted in Bitcoin market, where researchers found that the correlations between the sentiment information appeared in tweets and price movements (Kraaijeveld and De Smedt 2020; Pant et al. 2018). Meanwhile, influences towards the EMH generated from volumes of relevant tweets in Bitcoin market have also been well studied (Shen et al. 2019; Choi 2021), suggesting that the number of previous day tweets can significantly predict future RV and trading volume of Bitcoin. Additionally, different attributes of Twitter, such as number of replies, number of retweets, number of followers, etc., have been shown to have various impacts on both organization performance (Wang et al. 2017; Jung and Jeong 2020) and finance markets (Zhang et al. 2011; Otabek and Choi 2022; Nofer and Hinz 2015). Researchers, such as Jain et al. (2018); Aggarwal et al. (2019), have also combined Twitter attributes with sentiment analysis and found different correlations between interactive effects and price movements in Bitcoin market. For example, Mohapatra et al. (2019) considered that number of likes, retweets and followers play an important role during the prediction, while Otabek and Choi (2022) concluded that the influence of the number of followers is the most significant.

Thus, previous work has brought out a general idea about the efficiency of Bitcoin market through bridging price movements and meaningful information. However, both the sentiment and volume could only represent limited parts of properties of public information. Considering rhetorical modes as a more common classification for communication, words per se could be addressed to identify efficient/inefficient market effects in Bitcoin. Additionally, elimination of Twitter "bot" accounts should also be taken into account into the research, since they were found to contribute at least 1–14% of contents (Kraaijeveld and De Smedt 2020). Finally, while approving the EMH, it is necessary to take into consideration 24/7 high trading frequencies of Bitcoin featured by decentralized trading patterns as well as the time period after Sept 2017 when most of the fluctuations took place. We will scrutinize the aforementioned issues in our analysis.

### 3. Research Framework

It takes four steps to realize our research approach, each using different subsets of tweets. First, we extract semantic vector spaces of words from each training subset defined according to three different time intervals, namely hourly, 4-hourly, and daily, and confirm significant dissimilarities of information revealed between bullish and bearish markets. Second, we extract particular keywords and their relevant scores from different training sets. These keywords are used to identify whether the successive trading intervals will hold positive or negative returns. Third, we develop a mechanism to encode each tweet in line with keywords and relevant scores and finalize datasets for later classification models. In the final step, we feed encoded datasets to 6 classifiers and assess if the information retained in tweets from each time interval has predictive power over the market trend in the next trading interval.

There are five major components included in research methods, namely data collection and processing, a Global Vector (GloVe) model for determining semantic vector spaces of tweets towards different market movements, a text feature based automatic keyword extraction model (YAKE), feature vectors encoding mechanism, and Light Gradient Boosting Machine (LGBM) based classification models.

The proposed research framework is expected to be able to identify different groups of informative words in tweets collected ahead of different market movements, namely increasing and decreasing. If so, this would suggest that information carried in tweets will influence price changes in the market, expressly the market efficiency.

#### 3.1. Data Collection and Processing

Two sets of data are used in our research: market data collected from Gemini Exchange (gemini.com) and tweets. Unlike stocks and commodities, trading on crypto market is not regulated and lasts 24 hours a day across a decentralized network of computers, signaling that a significant stochastic volatility could occur in a shorter amount of time, such as an hour. To address the high frequency of price movements in Bitcoin market, we use hourly, 4 hourly, and daily time series as modeled variables and calculate the log-return of Bitcoin pricing data paired with US Dollars in different time intervals. The targeted period in our research spans from Sep 1 2017 to Sep 1 2022, because majority of significant Bitcoin fluctuations took place after Sep 2017 (Edwards 2022).

Even though other exchange platforms, such as Binance, Coinbase, Kraken, and Bitstamp, presently have larger trading volumes, Gemini Exchange is selected as the market data source due to several advantages. Gemini Exchange holds higher security standards and is notable for never having been subject to a major hack (Birken 2023), leading to a reliable historical database. Researchers (Giudici and Abu-Hashish 2019; Hu et al. 2022; Dyhrberg et al. 2018) have been applying Gemini Exchange data for different purposes in Bitcoin studies, since it is one of the earliest exchange platforms and focuses on trading only popular assets. Our research requires high-frequency data (hourly) with a long historical time span (5 years) which raises an issue about data availability from other sources. Moreover, we observed that general trends of market movements (increase/decrease) reported from different sources have little variance among themselves. Thus, since detailed price differences are not within our scope of work, the market data sources should not have a significant impact on the results of our study.

Twitter hashtags "$\#Bitcoin$" and "$\#BTC$" are used in our search query as preserved keywords to select most relevant tweets to Bitcoin topic. In order to filter out the noise we exclude blocked tweets, tweets from suspended accounts, tweets as promoted content, and tweets posted by bots. We also eliminate quoted tweets, retweets and reply tweets in order to keep only the original content. The research approach requires tweets posted one time interval ahead of the predicted price movement. Hence, a set of 28,739,514 English tweets with above criteria posted between Aug 31 2017 to Aug 31 2022 is retrieved using recursive calls to the Twitter API (v2.0). After converting tweets into lower case, we remove numbers, punctuations, emoticons, URLs, e-mail addresses, symbols, Unicode characters, and non-ASCII words. Three corpora of stop words, the Natural Language Toolkit (NLTK) Stopwords corpus (Bird et al. 2009), the Gensim Library (Rehurek and Sojka 2010), and a list of common prepositions and conjunctions, are implemented to further remove uninformative words (Polyzos and Wang 2022). We also removed query keywords "Bitcoin" and "BTC" to limit their disturbances to later applications of keywords extraction models. Stemming and lemmatization are not involved in data sanitization in order to keep the semantic meaning of tweet sentences and to increase sensitivities of models while operating on large amount of text vectors.

Based on each time interval and its consequent price movement (increase or decrease) in the market, we group tweets and generate six datasets of tweets, where we will operate analysis on influences of keywords of tweets towards price change. We summarize names, descriptions and total amount of tweets of each dataset in Table 1.

*3.2. Distances of Semantic Vector Spaces of Information*

*3.2.1. GloVe Vector Space of Semantics of Words Model*

Global Vectors for Word Representation (GloVe) (Pennington et al. 2014) is an unsupervised machine learning model, aiming to encode semantic meanings of words into vector space by taking account of ratios of word-word co-occurrence probabilities. Based on the method, the similarity of documents constructed by meanings of vectors can be calculated. In our research, each document represents a group of tweets labeled with one of two possible market movements, either bullish or bearish. A set of words from each document is a corpus. We suppose $m$ and $n$ are randomly selected target words in the corpus. The window size $u$ is predefined as a number of context terms adjacent to target words on both sides. In the model, $\mathbf{W}_{mn}$ is the co-occurrence matrix that records the number of times that the word $n$ appears in the context of $m$ within the range of the window size $u$. For the word $m$, $\mathbf{W}_m = \Sigma_t \mathbf{W}_{mt}$ counts the total number of times of any context word $t$ appearing next to $m$ within the window size $u$. We define $\Xi_{m,n}$ as the probability function that the word $n$ appears next to the word $m$, where $\Xi_{m,n} = \Xi(n|m) = \mathbf{W}_{mn}\mathbf{W}_m^{-1}$. The ratio of co-occurrence probability is expressed as

$$\Upsilon(\nu_m, \nu_n, \tilde{\nu}_t) = \frac{\Xi_{mt}}{\Xi_{nt}}, \qquad (1)$$

where $\tilde{\nu}_t \in \mathbb{R}^d$ are context vectors and $\nu_n, \nu_m \in \mathbb{R}^d$ are word vectors. After taking vector differences, the information contained in the $\mathbf{W}_{mt}(\mathbf{W}_{nt})^{-1}$ is encoded in the vector space. To avoid complication due to the linear structure during the machine learning training procedure, the dot product is implemented and Eq. (1) takes the form,

$$\Upsilon\left((\nu_m - \nu_n)^\intercal \tilde{\nu}_t\right) = \frac{\Xi_{mt}}{\Xi_{nt}}. \qquad (2)$$

Table 1: Description of Twitter Datasets

| DataSet | Description | Total Tweets |
|---|---|---|
| *DailyIn* | Tweets are posted during the previous day as positive returns appear in the following day. | 14,845,855 |
| *DailyDe* | Tweets are posted during the previous day as negative returns appear in the following day. | 13,886,765 |
| *4HourDe* | Tweets are posted during the previous four hours as positive returns appear in the following four hours. | 14,561,236 |
| *4HourIn* | Tweets are posted during the previous four hours as negative returns appear in the following four hours. | 14,171,384 |
| *HourIn* | Tweets are posted during the previous one hour as positive returns appear in the following hour. | 14,525,726 |
| *HourDe* | Tweets are posted during the previous one hour as negative returns appear in the following hour. | 14,177,494 |

The identities of $m$, $n$ and $t$ are interchangeable during the training procedure, leading to the following two conditions: $\nu \leftrightarrow \tilde{\nu}$ and $\mathbf{W} \leftrightarrow \mathbf{W}^\intercal$. To symmetrize Eq. (2) and retain the consistency of the model, $\Upsilon$ has to be a homomorphism between $(\mathbb{R}, +)$ and $(\mathbb{R}_{>0}, \times)$, such that

$$\Upsilon\left((\nu_m - \nu_n)^\intercal \tilde{\nu}_t\right) = \Upsilon(\nu_m^\intercal \tilde{\nu}_t) - \Upsilon(\nu_n^\intercal \tilde{\nu}_t) = \frac{\Upsilon(\nu_m^\intercal \tilde{\nu}_t)}{\Upsilon(\nu_n^\intercal \tilde{\nu}_t)}. \qquad (3)$$

Together with Eq. (2), Eq. (3) can be solved by substitution

$$\Upsilon(\nu_m^\intercal \tilde{\nu}_t) = \Xi_{mt} = \frac{\mathbf{W}_{mt}}{\mathbf{W}_m}. \tag{4}$$

The result of Eq. (3) is (Pennington et al. 2014)

$$\Upsilon = \exp \quad \text{or} \quad \nu_m^\intercal \tilde{\nu}_t = \log(\Xi_{mt}) = \log(\mathbf{W}_{mt}) - \log(\mathbf{W}_m) \quad . \tag{5}$$

Bias terms for both vectors should be included to finally symmetrize the model:

$$\nu_m^\intercal \tilde{\nu}_t + \epsilon_m + \tilde{\epsilon}_t = \log(\mathbf{W}_{mt}). \tag{6}$$

We treat the Eq. (6) as a least squares problem, bringing up a weighting function $f(\mathbf{W}_{mn})$ in to the cost function $Jx$, given by (Pennington et al. 2014)

$$Jx = \sum_{m,n=1}^{\varpi} f(\mathbf{W}_{mn})(\nu_m^\intercal \tilde{\nu}_n + \epsilon_m + \tilde{\epsilon}_n - \log(\mathbf{W}_{mn})^2, \tag{7}$$

where $\varpi$ is the total number of words. Here, $f(\mathbf{W}_{mn})$ has to satisfy two constraints. Namely, (1) $f(0) = 0$ and $\lim_{\alpha \to 0} f(\alpha) \log^2 \alpha < \infty$; (2) to rule out overweightness in case of rare or frequent co-existences, $f(\alpha)$ is not decrementable and has to be relatively small for large $\alpha$. In such cases, we conclude

$$f(\alpha) = \begin{cases} \alpha^\beta \alpha_{\max}^{-\beta} & \text{if} \quad \alpha < \alpha_{max} \\ 1 & \text{if} \quad else \end{cases} . \tag{8}$$

The model is pretrained on twitter data containing about $2 \times 10^9$ tweets, $27 \times 10^9$ tokens, and $1.2 \times 10^6$ words vocabulary (Pennington et al. 2014). Each vector is taken to be 200-dimensional. The parameters of our final tuned best performance model are $u = 10$, $\alpha_{max} = 100$, and $\beta = 3/4$, stochastically sampling nonzero elements from $\mathbf{W}$ with initial $learning\_rate = 0.05$, and 50 iterations. The model will generates two sets of vectors, $V$ and $\tilde{V}$, from which we obtain the final word vectors, $V + \tilde{V}$.

*3.2.2. Similarity Measurement*

Similarities of semantic vector space of words can be measured by cosine distance. In our research, cosine similarity metric was implemented for testing the semantic similarity of sets of tweets corresponding to probabilities of opposite market movements. Here, cosine similarity is defined as

$$\cos(\vec{\vartheta}_a, \vec{\vartheta}_b) = \frac{\vec{\vartheta}_a \cdot \vec{\vartheta}_b}{\parallel \vec{\vartheta}_a \parallel \parallel \vec{\vartheta}_b \parallel}, \tag{9}$$

where $\vec{\vartheta}_a$ and $\vec{\vartheta}_b$ are the sets of tweets labeled under different market movements, where words are pre-processed in vector spaces calculated through GloVe model.

*3.3. Keywords Extraction*

In our study, Yet Another Keywords Extraction Model (YAKE) (Campos et al. 2020), is adopted and implemented to extract keywords from tweets. In the model, multiple local text features are statistically extracted from a single document, leading to an automatic selection of the most relevant keywords of the document. The algorithm is applied in several steps described as follows.

We assume each dataset is a document $d$, because each dataset is composed of a group of tweets $\tau$ indicating one market signal, either an increase or a decrease. For each document $d_k$, where $d_k = (\tau_1, ..., \tau_J)$, $J$ is the total number of tweets $\tau_j$, and where for each tweet $\tau_j$, $j \in (1, ..., J)$. Here, $d_k$ contains $I$ words, which are tokenized as a corpus $C$ of terms $c_i$. Thus, $C_{d_k} = (c_1, ...c_I)$, for each $c_i$, $i \in (1, .., I)$. Moreover, $d_k$ is a member document of $D$, where $D = (d_1, ..., d_K)$ with $d_k \in D$ and $k \in (1, .., K)$. For each term, we extract a set of four features, which will be modeled to capture characteristics of individual term, specifically, (1) term position $cp$; (2) term frequency $cf$; (3) term relevancy $cr$; and (4) term dispersion $cd$.

**Term position** measures how early the term $c_i$ appears in the document $d_k$. Our assumption is that the terms that occur in the early tweets should carry a higher value than the terms that appear later (Campos et al. 2020), because early identification of new information can provide substantial profits in the market (Dimson and Mussavian 1998). Given $c_i \in \tau_j$ and $\tau_j \in d_k$, we take $p^{d_k}_{(\tau_j, c_i)}$ to be the position of the tweet $\tau_j$ in the document $d_k$ (starting from 0), where $p \in \mathbb{N}_0$ for any $d_k, \tau_j$ and $c_i$. Since the term $c_i$ is unlikely to occur only once, we introduce $P_{d_k} = \{p^{d_k}_{(\tau_j, c_i)}\}$ as a set of positions of different tweets in the document. The weight $G$ of the term position $cp$ of term $c_i$ in the document $d_k$ is calculated as:

$$G^{cp}_{d_k, c_i} = \ln\left[\ln(e + \mathbb{M}(P_{d_k}))\right] \tag{10}$$

where $\mathbb{M}(P)$ is the median function of $P_{d_k}$. Here, Euler's number $e$ is added to distinguish $p = \varnothing$ from $p = 0$, in cases where only the first tweet contains certain terms. Double $ln$ serves as a smoothing function to reduce the differences within a large number of different terms in a document. Given in this form, the weight $G^{cp}_{d_k, c_i}$ becomes smaller as the more terms $c_i$ appear in earlier positions in their corresponding documents.

**Term frequency** is a universal measurement of word significance (Luhn 1958) in a document. For each term $c_i$, its frequency in the document $d_k$ is $\omega_{d_k, c_i}$, where $\omega \in \mathbb{N}_0$ for any $c_i$ and $d_k$. In order to prevent a bias towards high frequency of term $c_i$ in a long text document $d_k$, the weight function $G$ of term frequency $cf$ is normalized as

$$G^{cf}_{d_k, c_i} = \omega_{d_k, c_i} \left(\frac{1}{I} \sum_{i=0}^{I} \omega_{d_k, c_i} + \sigma\right)^{-1}, \tag{11}$$

where $\sigma$ is the standard deviation of the set of $\omega_{d_k, c_i}$.

**Term relevancy** aims to capture the extent of relevancy of a term in its specific context. The smaller the number of different terms that coexist on both sides of the term $c_i$, the more important the term is (Machado et al. 2009). For each term $c_i$, let the window size $O$ be the number of terms adjacent to $c_i$ on both sides, where the terms $l_o$ and $r_o$ are on left and right side of it, respectively, with $o \in (1, ..., O)$. For the document $d_k$, we further define $\Psi_{d_k,c_i}(O)$ as the total number of *different* terms adjacent to the term $c_i$ and within the range of the window $O$, and $\varphi(l_o|r_o)$ as the number of coexisting terms $c_i$ with $l_o$ or $r_o$. The weight $G^{cr}_{d_k,c_i}$ of the term relevancy $cr$ of the term $c_i$ in the document is given by

$$G^{cr}_{d_k,c_i} = 1 + \left( \frac{\Psi_{d_k,c_i}(O)}{\sum \varphi_{d_k,c_i}(l_o)} + \frac{\Psi_{d_k,c_i}(O)}{\sum \varphi_{d_k,c_i}(r_o)} \right) \omega_{d_k,c_i} \left( \max_{i \in (0,I)} \omega_{d_k,c_i} \right)^{-1}, \qquad (12)$$

where $\omega_{d_k,c_i} \left( \max_{i \in (0,I)} \omega_{d_k,c_i} \right)^{-1}$ is a penalization function introduced to balance the weights of very common terms, such as "accept", which yield significantly higher value of $\omega_{d_k,c_i}$ and $\Psi_{d_k,c_i}(O)$ than less common English terms. Thus, low values of the weighting function $G^{cr}_{d_k,c_i}$ imply high term relevancy.

**Term dispersion** quantifies the frequencies of term $c_i$ appearing in different tweets $\tau_j$ in document $d_k$. A term with larger number of appearances in different tweets has a higher probability of being important (Campos et al. 2020). In the document $d_k$, the weight $G^{cd}_{d_k,c_i}$ of the term dispersion $cd$ is computed as

$$G^{cd}_{d_k,c_i} = \Phi_{d_k,c_i}(J)^{-1}, \qquad (13)$$

where $\Phi_{d_k,c_i}$ is the number of tweets in which $c_i$ appears.

Combining weights of all four features introduced above together, the keyword score $\kappa$ of the term $c_i$ in the document $d_k$ can be expressed as:

$$\kappa_{d_k,c_i} = G^{cp}_{d_k,c_i} G^{cr}_{d_k,c_i} \left( \frac{G^{cf}_{d_k,c_i}}{G^{cr}_{d_k,c_i}} + \frac{G^{cd}_{d_k,c_i}}{G^{cr}_{d_k,c_i}} \right)^{-1}. \qquad (14)$$

Here, the divisor $G^{cr}_{d_k,c_i}$ mitigates the high values that occur when the term $c_i$ with high term frequency is present in a large volume of different tweets even though it is not relevant. In other words, it is an important indication of term significance when both $G^{cd}_{d_k,c_i}$ and $G^{cf}_{d_k,c_i}$ become large when $G^{cr}_{d_k,c_i}$ is small (*i.e.*, more relevant). Likewise, the factor $G^{cr}_{d_k,c_i}$ strengthens the relative weight $G^{cp}_{d_k,c_i}$ in cases when the term $c_i$ appears at the beginning or as one of the first few words in a tweet, potentially indicating importance, but has a low weight $G^{cr}_{d_k,c_i}$ because there are not many terms on its left side. In general, a term with a lower $\kappa_{d_k,c_i}$ value is more significant and more likely to be a keyword.

**Term deduplication** aims to discard similar terms after obtaining a list of ranked terms based on their keyword score. Ratcliff-Obershelp similarity algorithm (Ratcliff and Metzener 1988) technique is implemented to finalize the list of keywords from ranked terms. The similarity score $r$ is calculated as $r = 2\Pi(\Gamma)^{-1}$, where $\Pi$ operator sums the lengths of all matched sequences with different patterns in given two terms, and $\Gamma$ is the total length of two strings (terms). We feed the Ratcliff-Obershelp similarity function into the loop of paired terms comparison with a predefined threshold $\theta$. To assure the uniqueness of individual words from a large volume of data we set $\theta = 0.9$ to maintain a strict dissimilarity threshold. In case of $r \geqslant \theta$, we discard the term that has a higher feature score. After this procedure, the final list of keywords will be extracted in ascending order of keyword score.

***Term Unbiasedness*** is a measure used in keywords extraction model to assess the unavoidable randomness during the final keywords selection. It consists of three contributions. First, the unpredictable property of user-generated content on social media (Twitter) determines that the presence and the linguistic meaning of each posted term are random. When selecting the final keywords, we measured four features for each individual term: (1) term position $cp$; (2) term frequency $cf$; (3) term relevancy $cr$; and (4) term dispersion $cd$, all according to different relationships between the term and documents, but not between the linguistic meaning of each term and documents. Second, the keywords extraction model aims to select the term in the vocabulary within a very large amount of random (i.e., mutually disconnected and uncorrelated) words. To assure that the extraction procedure is unbiased, we assign equal importance to each term regardless of the semantic roles, its format, and other perspectives in the document. Lemmatization, stemming and other methods with homologous purpose are specifically not applied to guarantee that each word with random formats (such as a suffix, a prefix, etc.) is treated equally and, consequently, have the same probability to be ranked and extracted. For instance, "expected" and "expecting" is considered as different terms to avoid any biases. Third, Bitcoin is by construction a "random" topic in linguistics. Keywords related to it would often look meaningless and random in general communication.

*3.4. Feature Vectors Encoding*

In order to feed the tweets into final classifier models, we have to encode each tweet with feature variables (keywords) and meaningful weighted feature (keyword) scores. We first take each extracted keyword and count the number of occurrences $\mathfrak{O}$ of each keyword $c_i$ in each tweet $\tau_j$. The importance of keywords of each set should be reflected in the corresponding set of tweets to preserve value of information, which might be lost if only counting occurrences. Note that the keyword scores derived from the YAKE model are arbitrary numbers with certain orders. To make these numbers more meaningful, for each keyword, we calculate the proportion of its score to the sum of keyword scores of all selected keywords $\widehat{C}_{d_k}$, where $\widehat{C}_{d_k} \subseteq C_{d_k}$. We then subtract it from 1 to correct for the inverse relation between the weights of importance and the keyword scores, and finally multiply it by one half to preserve the equal ratio of keywords corresponding to two possible market movements, either an increase or a decrease. Putting everything together, for each tweet, the weighted feature score of each feature (keyword) is encoded and expressed as

$$\mathring{\kappa}_{\tau_j, c_i} = \frac{1}{2} \mathfrak{O}_{c_i, \tau_j} \left( 1 - \frac{\kappa_{d_k, c_i}}{\sum_{c_i \in \widehat{C}_{d_k}} \kappa_{d_k, c_i}} \right). \qquad (15)$$

*3.5. Light Gradient Boosting Machine Classifiers*

In our research, we use Light gradient boosting machine (LGBM) (Ke et al. 2017), an efficient machine learning algorithm, to classify the relation between different market returns and tweets. The LGBM is based on Gradient Boosting Decision Tree (GBDT) (Friedman 2001) and implemented with Gradient-based One-Side Sampling (GOSS) and Exclusive Feature Bundling (EFB) to promise a fast and light version of the classifier algorithm capable of high level of accuracy.

*3.5.1. GBDT*

The baseline decision tree models of LGBM/GBDT leverages ensemble learning technique to generalize final prediction by combining predictions of multiple decision trees together. We assume that our model includes $Tr$ trees, where $tr$ is the $tr-th$ tree, and a training set $\mathcal{Z} \ni \{\varkappa_h\}$, where input sample $\varkappa_h$ is a $\varsigma$ dimension vector in space $\aleph^\varsigma$ and $H$ is the size of set $\mathcal{Z}$ and for each $h$, $h \in (1,...,H)$. The target value of each $\varkappa_h$ is $y_h$. Given $\varkappa_h$, the output of GBDT model is

$$F_{Tr}(\varkappa_h) = \sum_{tr=1}^{Tr} \hat{y}_{(tr,h)}, \hat{y}_{(tr,h)} = \beth_{tr}(\varkappa_h), \tag{16}$$

where $f_{tr}(\varkappa_h)$ is the learning function for the $tr-th$ decision tree. The learning procedure of GBDT is from the input space $\aleph^\varsigma$ to the gradient space $\Delta$ (Ke et al. 2017). An additive training process is introduced into each new tree model for predicting gradients of the prior tree models. In other words, new boosting models are applied in each iteration of gradient descent. Thus, it takes several steps to obtain Eq. 16. Initially, we set $\hat{y}_{(tr=0,h)} = F_0(\varkappa_h) = 0$. Corresponding to outputs during each iteration of gradient boosting, the negative gradients of the loss function takes the form: $\delta_{(tr,h)} = y_h - \hat{y}_{(tr-1,h)} = y_h - F_{tr-1}(\varkappa_h)$. We feed $\{\delta_{(tr,h)}\}$ and $\{\varkappa_{(h)}\}$ into $\beth_{tr}(\varkappa_h)$ and update $F_{tr}(\varkappa_h) = F_{tr-1}(\varkappa_h) + \hat{y}_{(tr,h)} = F_{tr-1}(\varkappa_h) + \beth_{tr}(\varkappa_h)$. Finally, after iterating $tr$ from 1 to $Tr$, $Tr$ decision trees are generated, leading to identical predictions of Eq. 16:

$$F_{Tr}(\varkappa_h) = F_{Tr-1}(\varkappa_h) + \hat{y}_{(Tr,h)} = \sum_{tr=1}^{Tr} \hat{y}_{(tr,h)}, \hat{y}_{(tr,h)} = \beth_{tr}(\varkappa_h). \tag{17}$$

To establish $\beth_{tr}(\varkappa_h)$, we need to select the splitting feature $f$ and optimal splitting point $\zeta$. In our case, each feature is represented by a keyword. The procedure of selection will be discussed in the following section. After each splitting, minimization of the square error between the predicted value and the real value of children nodes should be achieved to get $\beth_{tr,f,\zeta}(\varkappa_h)$, defined as (Friedman 2001)

$$\beth_{tr,f,\zeta}(\varkappa_h) = \min_{f,\zeta}\left[\min_{\mathfrak{x}_{(l,tr,f,\zeta)}} \sum_{\substack{\varkappa_h \in \mathcal{Z} \\ \varkappa_{h,f} \leqslant \zeta}} (\delta_{tr,h} - \mathfrak{x}_l)^2 + \min_{\mathfrak{x}_{(r,tr,f,\zeta)}} \sum_{\substack{\varkappa_h \in \mathcal{Z} \\ \varkappa_{h,f} > \zeta}} (\delta_{tr,h} - \mathfrak{x}_r)^2 \right], \tag{18}$$

where $\mathfrak{x}_{(l|r,tr,f,\zeta)}$ are the predicted values of left and right node after splitting.

*3.5.2. Information Gain and GOSS*

One splitting criteria of decision tree is to maximize information gain through splitting each node at the feature that leads to purest child nodes. In GBDT, using the training set described above, for a fixed node, the information gain $\daleth$ of feature $f$ split at point $\zeta$ can be expressed as (Friedman 2001)

$$\daleth_{f|\mathcal{Z}}(\zeta) = \frac{1}{H}\left(\left(\sum_{\substack{\varkappa_h \in \mathcal{Z} \\ \varkappa_{h,f} \leqslant \zeta}} \delta_h\right)^2 (H^f_{l|\mathcal{Z}}(\zeta))^{-1} + \left(\sum_{\substack{\varkappa_h \in \mathcal{Z} \\ \varkappa_{h,f} > \zeta}} \delta_h\right)^2 (H^f_{r|\mathcal{Z}}(\zeta))^{-1}\right), \tag{19}$$

where $H^f_{l|\mathcal{Z}}(\zeta) = \| \{\varkappa_h \in \mathcal{Z} | \varkappa_{h,f} \leqslant \zeta\} \|$ and $H^f_{r|\mathcal{Z}}(\zeta) = \| \{\varkappa_h \in \mathcal{Z} | \varkappa_{h,f} > \zeta\} \|$; here $r$ and $l$ are the right and left child nodes, respectively. The maximum information gain of feature $f$ is selected at point $\check{\zeta}$, where

$$\beth_f(\check{\zeta}) = \underset{\zeta}{\operatorname{argmax}} \beth_f(\zeta).$$

Note that we omit the index $\mathcal{Z}$ in notations for simplicity.

GOSS method advances GBDT through weighing contributions of larger gradients towards information gain. Let $lg$ and $sg$ be the sampling ratio of instances with large and small absolute values of gradients. Top $\check{lg} = lg \times H$ data are selected and form a subset $Lg$. Subset $Lg^{\complement}$ contains remaining $H - \check{lg}$ data, where number $\check{sg} = sg \times \| Lg^{\complement} \|$ samples are randomly drawn out and construct a subset $Sg$. The information gain $\tilde{\beth}_f(\zeta)$ over $Lg \bigcup Sg$ is updated as (Ke et al. 2017)

$$\tilde{\beth}_f(\zeta) = \frac{1}{H} \left[ \left( \sum_{\varkappa_h \in Lg_l} \delta_h + \frac{1-lg}{sg} \sum_{\varkappa_h \in Sg_l} \delta_h \right)^2 (H^f_l(\zeta))^{-1} + \left( \sum_{\varkappa_h \in Lg_r} \delta_h + \frac{1-lg}{sg} \sum_{\varkappa_h \in Sg_r} \delta_h \right)^2 (H^f_r(\zeta))^{-1} \right], \tag{20}$$

where $\frac{1-lg}{sg}$ serves for normalization purposes and $Lg_l = \{\varkappa_h \in Lg | \varkappa_{h,f} \leqslant \zeta\}$, $Lg_r = \{\varkappa_h \in Lg | \varkappa_{h,f} > \zeta\}$, $Sg_l = \{\varkappa_h \in Sg | \varkappa_{h,f} \leqslant \zeta\}$, $Sg_r = \{\varkappa_h \in Sg | \varkappa_{h,f} > \zeta\}$. The error terms can be derived as

$$\mathcal{E}(\zeta) = |\tilde{\beth}_f(\zeta) - \beth_f(\zeta)|$$

$$\bar{\delta}^f_l(\zeta) = \left( \sum_{\varkappa_h \in (Lg \bigcup Lg^{\complement})_l} |\delta_h| \right) (H^f_l(\zeta))^{-1}$$

$$\bar{\delta}^f_r(\zeta) = \left( \sum_{\varkappa_h \in (Lg \bigcup Lg^{\complement})_r} |\delta_h| \right) (H^f_r(\zeta))^{-1}.$$

We denote $E_{lg,sg} = (1-lg)(\sqrt{sg})^{-1} \max_{\varkappa_h \in Lg^{\complement}} |\delta_h|$ and $F = \max(\bar{\delta}^f_l(\zeta), \bar{\delta}^f_r(\zeta))$. With the probability equal to at least $1 - \rho$, we can get

$$\mathcal{E} \leqslant E^2_{lg,sg} \ln(\rho^{-1}) \max\left\{ \frac{1}{H^f_l(\zeta)}, \frac{1}{H^f_r(\zeta)} \right\} + 2F E_{lg,sg} \sqrt{\frac{\ln(\rho^{-1})}{H}}, \tag{21}$$

By measuring the difference of the information gain calculated by the sampling method in GOSS and the original distribution, the generalization error in GOSS takes form as $\mathcal{E}^{GOSS}_{gen}(\zeta) = |\beth_f(\zeta) - \beth_*(\zeta)|$, where $\mathcal{E}^{GOSS}_{gen}(\zeta) \leqslant |\tilde{\beth}_f(\zeta) - \beth_f(\zeta)| + |\beth_f(\zeta) - \beth_*(\zeta)| \triangleq \mathcal{E}_{GOSS}(\zeta) + \mathcal{E}_{gen}(\zeta)$ (Ke et al. 2017). Hence, the error in GOSS will be very close to the error calculated on the full dataset, even though GOSS will incur significantly reduced computation costs due to the applied sampling procedure. Moreover, after the sampling procedures, the gradients corresponding to GBDT are changed to $\{\frac{1-lg}{sg} \times \delta_h\}$.

*3.5.3. EFB*

Exclusive feature bundling(EFB) is introduced in LGBM to speed up the training process through bundling exclusive features into a single one. A greed bundling based algorithm (Ke et al. 2017) is deployed to decide which features should be bundled together. The procedure can be qualitatively described in three steps. First, a graph coloring problem is defined, providing vertices as features and edges for every two vertices if represented features are not mutually exclusive. Each edge is given a weight proportional to the number of conflicts (overlapped nonzero values between features). Second, given the number of incoming edges as degree of conflicts, features (vertices) are ranked in descending order. Finally, after looping through the ranking list, each feature is either assigned to an existing bundle (if degree is below predefined max conflict rate) or to a newly created bundle.

To merge the features into a bundled feature the algorithm extends the range (bin) of value of every original feature to include all offsets and assign the value according to the new range. For example, assume that the sample $\varkappa_{1,\ldots,5}$ on features A and B have value bins $[0, 5)$ and $[0, 10)$, i.e. $\varkappa_{1,\ldots,5}^{A} \in [0, 5)$ and $\varkappa_{1,\ldots,5}^{B} \in [0, 10)$. Let the bundled feature C has an extended range to include both A and B. To do so, A still holds the bin $[0, 5)$, but B is translated to the bin $[5, 15)$. If $\varkappa_{1,\ldots,5}^{(A,B)} = \{(0, 2), (0, 8), (4, 0), (1, 0), (0, 0)\}$, then the new bundled feature $\varkappa_{1,\ldots,5}^{(C)} = \{7, 13, 4, 1, 0\}$. Therefore, unnecessary computations for zero-feature values is avoided. Moreover, exclusive continuous features are packed into reduced numbers of discrete bins, accelerating the training procedure by reducing the total size of features.

### 3.6. Methodology Pipeline

Python language and libraries GloVe (Pennington et al. 2014), YAKE (Campos et al. 2020) and LightGBM (Ke et al. 2017) are used in our research for measuring semantic vector spaces, extracting keywords and feature scores, and finalizing classifiers, respectively.

As the first step to test the market efficiency hypothesis, we treat all information during each market movement as a whole and implement similarity metrics on vector spaces of information. We are aware that large amount of information noise could appear during nearly flat periods and will disturb the model accuracy, because of tremendously large information volume in the research. We assume the information should indicate strongest signals at highest and lowest points of market returns. Information extracted during these times could result in more meaningful predictions towards market efficiency. Thus, we first cluster tweets posted during each time interval ahead when markets show the largest increase or decrease and then derive the similarity score and keywords from these clusters for the next step operating. In order to filter biases from overlaps of unavoidable information between opposite market movements, the dissimilarity of semantic vector spaces will be implemented as a baseline weight in final classifiers models.

After applying YAKE model, we obtain 6 sets of ranked keywords and their scores according to opposite market movements and time intervals that predefined in previous section. We compare every two sets within the same interval and take the first $64=2^8$ most important and unique keywords from each set as feature variables. Considering the tradeoff among our large volume of data, computational habit and costs, and model accuracy with increased number of feature variables, 64 selected as the number of feature variables is not completely arbitrary. We encode these features/keywords in the corresponding dataset and also in the dataset where opposite market movement within same time interval is held. Because keywords will also show up as random information during contradictory movements. In such cases, occurrences of keywords should be taken account as a proxy for reverse validation of market efficiency. In other words, key information is expected to bring out high probabilities or positive relations towards predictions of expected movements, as well as low probabilities or negative relations towards predictions of contradictory movements. Therefore, according to 6 sets of keywords, we encoded 6 datasets of tweets to 12 sets, reorganized them, and obtained 6 final datasets. Furthermore, we label the positive and negative market movements as "1" and "0", respectively. We summarize this procedure in Table 2.

Table 2: Features Encoded Datasets.

| Keywords | Interval | Tweets Sets | | Encoded |
|---|---|---|---|---|
| Increase | Hourly | *HourIn* | *HourDe* | *HkIn* |
| Decrease | | | | *HkDe* |
| Increase | 4-Hourly | *4HourIn* | *4HourDe* | *4HkIn* |
| Decrease | | | | *4HkDe* |
| Increase | Daily | *DailyIn* | *DailyDe* | *DkIn* |
| Decrease | | | | *DkDe* |

\* The encoded data set includes both sets of tweets within the same time interval

Before fitting LGBM models, we apportion each encoded dataset into training and testing sets, with a ratio of 80-20 split. We deploy GOSS and EFB in LGBMs for improving efficiency of our proposed predictive models. We set the number of leaves to 31 and the minimal number of data points in one leaf to 20 in order to avoid "overfitting" in LGBM. We do not limit the maximum depth, due to the large data size. The learning rate is set to 0.01. With these parameters, 10-fold cross-validations with 80-20 training and validation split are used to assess the accuracy of the model on the training set. Through this procedure, LGBM models are finalized and fitted on testing sets to predict the probability of increase or decrease movements according to the information in prior times. This probability will be adjusted with the baseline weight calculated from vector spaces. Meanwhile, we treat the two possible market movements as a binary classification problem and explore the prediction accuracy of future market movements by providing certain information. We also propose two types of robustness tests to check the consistency of our results. As the first check, we apply the basic Gradient Boosting Decision Tree (GBDT) technique without GOSS and EFB tuning hyper-parameters to eliminate possible biases of results from different setup of models. In the second test, we keep the original setups of proposed predictive models, but change the split ratio of training and testing sets to 70-30 for assessing the reliability of models. The schematic representation of the procedure is given in Figure 1.

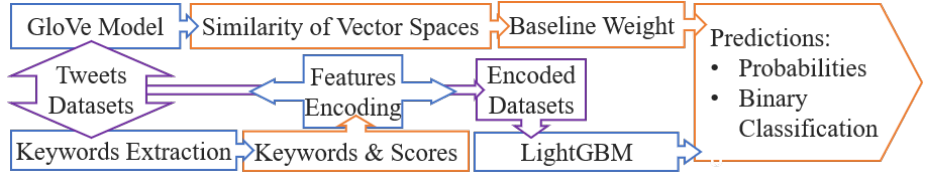

Figure 1: Schematic Representation of the Methodology Pipeline.

## 4. Empirical Results

After calculating distances of semantic vector spaces of all information extracted ahead of the most fluctuating market movements within different intervals, we notice that the information yielding the least significant difference between opposite movements fall in one hour duration, implying that the information in such a short time interval is likely to hold more noise that obfuscates the market changes. As a result, the initial reactions of the market appear to be weaker. Meanwhile, the information retrieved on a daily base more efficiently influence the market movements due to large accumulated dissimilarities between semantic vector spaces. Overall, to a certain extent, all available public information retrieved from Twitter with targeted topic, "Bitcoin", during different time intervals incorporate price changes in the market, indicating a preliminary conclusion of market efficiency. The sensitivity of efficient market reactions will be discussed further as the following models are applied. Table 3 summarizes similarities of semantic vector spaces of information on different interval basis. The smaller similarity number indicates the more distant information revealed ahead of bearish and bullish markets. The baseline weight is the distance, which will be included in subsequent calculations.

Table 3: Similarity between Vector Spaces of Information ahead of Increase and Decrease Movements.

| **Interval** | *Hourly* | *4 Hourly* | *Daily* |
|---|---|---|---|
| **Similarity** | 0.0249 | 0.0048 | 0.0034 |
| **Baseline Weight ($Base_{wt}$)** | 0.9750 | 0.9952 | 0.9965 |

We extract 6 sets of 64 keywords and illustrate them in word clouds in Figure 2. It is noticeable that some keywords retrieved from different intervals ahead of the same type of market movements are overlapped even though they do not hold same level of keyword scores and importance. This discovery implies that the possibility of information similarities happens across the time periods during which market movements are monotonic, either increasing or decreasing. However, we consider that the relations between market movements and contents of these words per se are arbitrary and thus fall outside of the scope of our research. These words serve purely as corridors of information-sharing that affect price determination. It is necessary to clarify the coincidental existence of certain proper nouns in the keywords, such as "Latoken", "China", etc. Even though these words likely have special meanings in cryptocurrency markets, analyses and comparison of the impacts of their representations should not be in line with the results of our study.

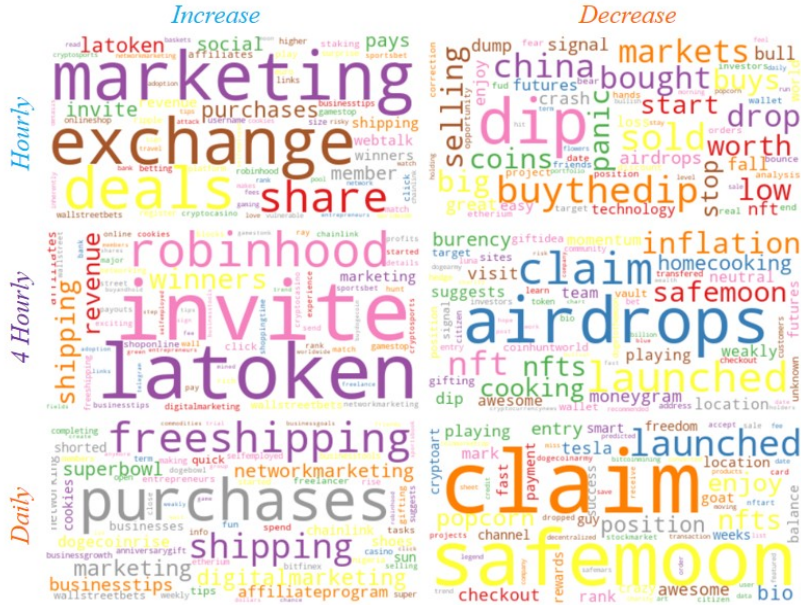

*Note: keywords in the left column represent key information available 1 hour, 4 hours, and 1 day ahead of the price increase and in the right column represent the same intervals ahead of the price decrease. In each word cloud map, the font size of keywords reflects relative levels of their importance in the corresponding dataset. Larger size keywords carry more weight in the set.*

Figure 2: Word Clouds of Keywords ahead of Different Market Movements and Time Intervals.

Following the feature encoding method, each tweet is encoded with feature scores of 64 variables and used to compose 6 new datasets, where LGBM models will be implemented. We address several important issues here. First, we are aiming to address whether the information in a certain interval could determine the succeeding market change, not the predictive power from a single tweet. Hence, we have to find a way to group feature scores together according to time intervals. In such a case, if we simply sum the scores of each variable in a single time interval, we would face the problem that the time intervals containing a larger volume of tweets would have larger summation. To avoid biases from uneven volume, our first group of baseline models takes the summation and divides it by the total number of tweets in the interval. This will give us a general idea about how original information influence market movements. Second, we have to take into account that not all tweets have the same reach. Information from tweets with more retweeted number may be more available for the public. In our second group of weighted models, products of feature scores from each tweet and its retweet number are added to the summation and then divided by the total number of tweets in the interval. Thus, the predicted influence on market movements is well adjusted by information diffusion. Our last group of weights-distributed classifiers accounts for the influence from average unit information propagated on Twitter in a certain interval, by dividing the summation from the second model with the total amount of retweets and tweets in the interval.

Each of these three LGBM classifiers is first validated on 6 datasets. In Figure 3

plots of learning curves of validation processes are reported. We notice that for each model, the accuracy and loss from both training and validation show improvements and an increase/decrease to a point of stability after certain iterations, implying a good fit of the model. Additionally, among the three different intervals, training procedures on daily base data appear to exhibit the best performances across different models. The best fitting models are in the baseline group, as evidenced by the smallest gaps between the respective validation and training plots. Overall, after 10-folds cross validations on 80% training data, we expect similar patterns while training and testing on entire datasets.

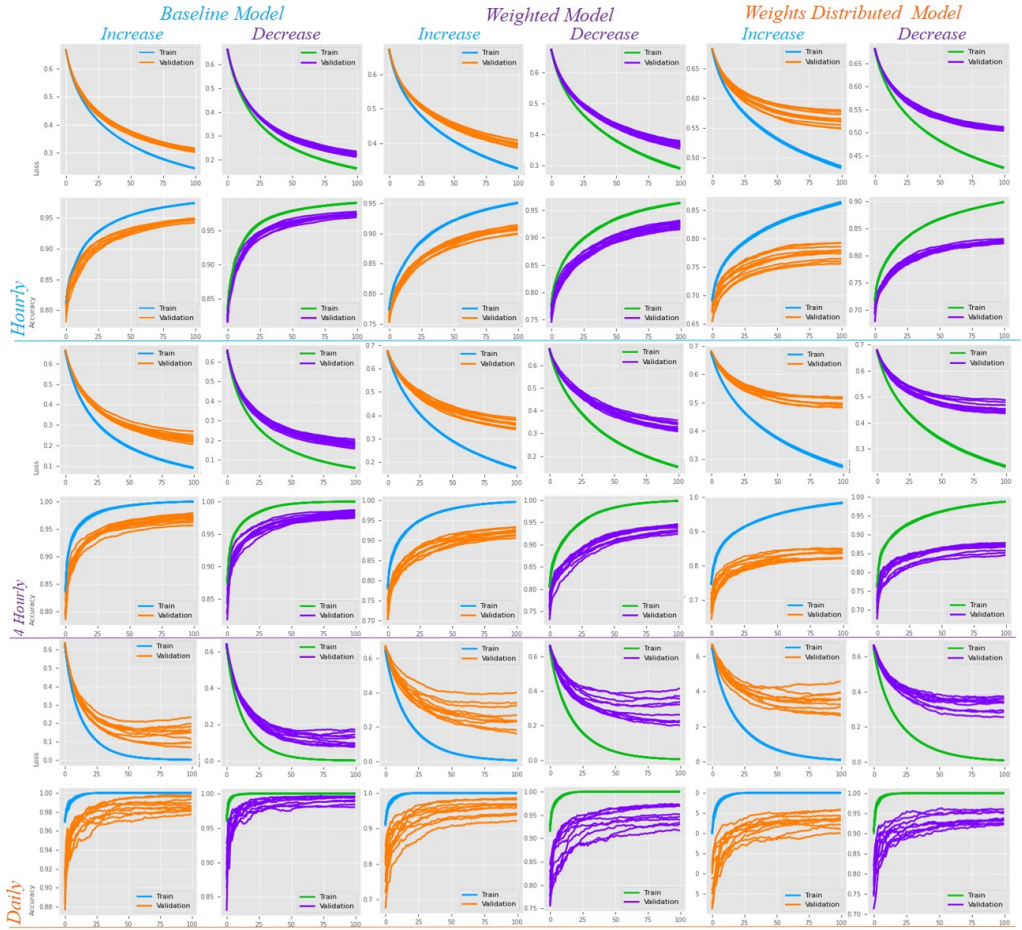

*Note: for each model, two plots are generated, namely loss (on the top) and accuracy (on the bottom). The x-axis of each plot is the number of iteration. Each model is corresponding to datasets from three time intervals. Increase and decrease denote the type of encoded features(keywords).*

Figure 3: Learning Curves (Accuracy and Loss) of 10 Folds Cross-Validations of LGBM.

By treating the increasing and decreasing trends as binary classes in LGBMs, we were able to assess the accuracy of predictions that could be made based on our feature variables. Results in the form of confusion matrices and performance metrics of each model are reported in Figure 4 and Table 4.

Table 4: Performance Metrics of LGBMs

| Interval Feature Set | Label | Baseline Model | | | Weighted Model | | | Weights Distributed Model | | | |
|---|---|---|---|---|---|---|---|---|---|---|---|
| | | Pre | Rec | F1 | Pre | Rec | F1 | Pre | Rec | F1 | Supt |
| **HkIn** | **DeCr** | 0.8761 | 0.8864 | 0.8812 | 0.8234 | 0.8127 | 0.8180 | 0.6843 | 0.7144 | 0.6990 | 4314 |
| | **InCr** | 0.8881 | 0.8779 | 0.8830 | 0.8199 | 0.8302 | 0.8250 | 0.7094 | 0.6790 | 0.6939 | 4430 |
| | *Acc* | | | **0.8821** | | | **0.8216** | | | **0.6965** | |
| **HkDe** | **DeCr** | 0.9230 | 0.9193 | 0.9211 | 0.8464 | 0.8403 | 0.8433 | 0.7575 | 0.7153 | 0.7358 | 4314 |
| | **InCr** | 0.9217 | 0.9253 | 0.9235 | 0.8456 | 0.8515 | 0.8485 | 0.7370 | 0.7770 | 0.7565 | 4430 |
| | *Acc* | | | **0.9223** | | | **0.8460** | | | **0.7466** | |
| **4HkIn** | **DeCr** | 0.9067 | 0.9299 | 0.9182 | 0.8319 | 0.8485 | 0.8401 | 0.7302 | 0.7434 | 0.7367 | 1056 |
| | **InCr** | 0.9330 | 0.9107 | 0.9217 | 0.8559 | 0.8400 | 0.8478 | 0.7563 | 0.7436 | 0.7499 | 1131 |
| | *Acc* | | | **0.9200** | | | **0.8441** | | | **0.7435** | |
| **4HkDe** | **DeCr** | 0.9503 | 0.9242 | 0.9371 | 0.8601 | 0.8561 | 0.8581 | 0.7502 | 0.7623 | 0.7562 | 1056 |
| | **InCr** | 0.9310 | 0.9549 | 0.9428 | 0.8662 | 0.8700 | 0.8681 | 0.7747 | 0.7630 | 0.7688 | 1131 |
| | *Acc* | | | **0.9401** | | | **0.8633** | | | **0.7627** | |
| **DkIn** | **DeCr** | 0.9632 | 0.9892 | 0.9760 | 0.9213 | 0.8865 | 0.9036 | 0.8492 | 0.8216 | 0.8352 | 185 |
| | **InCr** | 0.9886 | 0.9611 | 0.9746 | 0.8877 | 0.9222 | 0.9046 | 0.8226 | 0.8500 | 0.8361 | 180 |
| | *Acc* | | | **0.9753** | | | **0.9041** | | | **0.8356** | |
| **DkDe** | **DeCr** | 0.9890 | 0.9676 | 0.9781 | 0.8783 | 0.8973 | 0.8877 | 0.8939 | 0.8649 | 0.8791 | 185 |
| | **InCr** | 0.9674 | 0.9889 | 0.9780 | 0.8920 | 0.8722 | 0.8820 | 0.8656 | 0.8944 | 0.8798 | 180 |
| | *Acc* | | | **0.9781** | | | **0.8849** | | | **0.8795** | |

\* **Interval Feature Set**: the dataset with encoded features (keywords) of each interval, described in previous section.
\* **DeCr**, **InCr** labels that prices will increase/decrease in next time interval.
\* **Pre**: Precision; **Rec**: Recall; **Supt**: number of tested samples; **Acc**: Accuracy.

Final results are consistent with the ones obtained from cross validations. The highest accuracy from the baseline model indicates that the tweets with organic content have strongest reflection on market movements. Additionally, volatility trends tend to be corrected over a somewhat longer term with more effective information accumulated and diffused, as the accuracy score across the three groups of classifiers invariably favors predictions based on slightly longer periods. The predictions based on 1-hour windows yield the lowest accuracy while the ones on a daily basis yield the highest accuracy. We emphasize that our results show that, in most cases, the information related to decrease movements have better predictive accuracy than information related to increase movements. The differentiation is in range from 0.28% to 5.01%. The same is true in the comparison of precision and recall score between the two types of informative classifiers within the same time interval. However, we cannot identify whether an information-oriented predicting classifier would be more veracious towards bearish markets or bullish markets. Even in the same model, scores of precision and sensitivity(recall) towards distinguishing different movements cannot be homogeneous. Sometimes the precision of one score is higher than another one while recall scores show the opposite behavior. Thus, performances of LGBMs towards detecting different market movements are comparable. This finding very much resembles discussions in literature, with some papers concluding higher or similar model predictive power in bear markets (Coakley and Fuertes 2006; Chen 2009), whereas others obtained higher predictive power in bull markets (Kim and Nofsinger 2007).

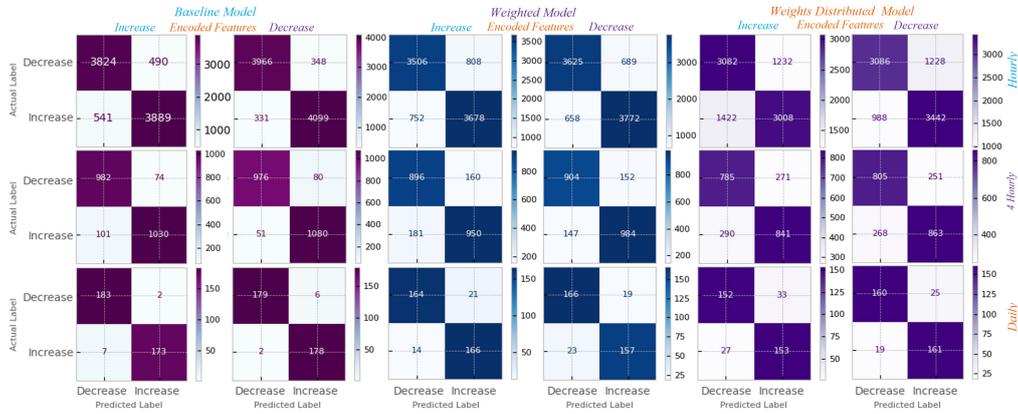

*Note: According to differences of predicted and actual labels in each model, 4 classes of results – true(false) increase(decrease) – are demonstrated. On the top line, increase and decrease denote the type of encoded features(keywords) corresponding to each classifier and a certain time interval.*

Figure 4: Confusion Matrices of 3 Groups of LGBMs.

The complete results of probability prediction for each classifier are given in Table 5. We also include the weights derived from distances of vector spaces between opposite movements to adjust the probabilities and assure that the final results are robust. Moreover, we report several interesting findings, all of which support the presence of market efficiency in Bitcoin market. As expected, models with *increase* labeled keywords yield predictions above 60% adjusted probability of positive market movement in the following time interval, and vice versa. Additionally, the keywords extracted from *decrease* data are showing slightly higher correctness when correlated with the price changes. We suspect that is related to the uniqueness of Bitcoin market history. Namely, the value of Bitcoin has kept rapidly increasing for a very long time from its inception, leading to a regular expectation of an overall positive movement. Hence, unexpected negative information will draw more attention frequently resulting in stronger reactions on the market. Furthermore, losses and disadvantages have greater impact on preferences than gains and advantages (Tversky and Kahneman 1991). Loss aversion for economic behaviors implies that investors react stronger while encountering negative shocks than positive shocks.

Table 5: Complete Results of Probability and Adjusted Probability of Market Movements Prediction by Proposed LGBMs with GOSS and EFB

| Interval Feature Set | Baseline Model | | | | Weighted Model | | | | Weights Distributed Model | | | |
|---|---|---|---|---|---|---|---|---|---|---|---|---|
| | $P_{Increase}$ | | $P_{Decrease}$ | | $P_{Increase}$ | | $P_{Decrease}$ | | $P_{Increase}$ | | $P_{Decrease}$ | |
| | *Prob* | *AdjP* | *Prob* | *AdjP* | *Prob* | *AdjP* | *Prob* | *AdjP* | *Prob* | *AdjP* | *Prob* | *AdjP* |
| *HkIn* | 80.07% | 78.06% | 20.84% | 20.32% | 74.11% | 72.26% | 28.25% | 27.54% | 61.96% | 60.42% | 39.41% | 38.42% |
| *HkDe* | 15.15% | 14.77% | 85.21% | 83.08% | 24.49% | 23.88% | 74.90% | 73.03% | 34.12% | 33.27% | 65.09% | 63.46% |
| *4HkIn* | 85.04% | 84.63% | 14.65% | 14.58% | 76.35% | 75.99% | 24.15% | 24.04% | 67.29% | 66.97% | 33.90% | 33.74% |
| *4HkDe* | 10.18% | 10.13% | 88.20% | 87.77% | 23.07% | 22.96% | 77.36% | 76.99% | 31.14% | 30.99% | 70.66% | 70.32% |
| *DkIn* | 94.36% | 94.03% | 4.52% | 4.51% | 86.80% | 86.50% | 17.59% | 17.53% | 81.34% | 81.06% | 22.02% | 21.95% |
| *DkDe* | 5.65% | 5.63% | 94.94% | 94.60% | 17.68% | 17.62% | 83.79% | 83.50% | 19.98% | 19.91% | 80.45% | 80.17% |

\* $P_{Increase}$, $P_{Decrease}$ is the probability that prices will increase/decrease in next time interval.
\* *Prob* is the original probability predicted from LGBMs; *AdjP* is the adjusted probability ($AdjP = Prob \times Base_{wt}$).
\* **Interval Feature Set** is the dataset with encoded features (keywords) of each interval, described in previous section. *In* and *De* indicate types of keywords (increase or decrease) being encoded and expected market movements being predicted.

In terms of higher probability and significant distances of keywords on a daily base model and prediction, we consider several possible causes. First, delayed market reactions (McQueen et al. 1996) do exist, so within certain thresholds of time, Bitcoin market may need a bit more time to exhibit a significant reflection on information. Second, the availability of public information plays an important role in the EMH, suggesting that the reach of effective information is insufficient in 1 hour, or in 4 hours, as compared to a day. Third, the information acquired on Twitter cannot provide equally trustworthy content as official announcements or reports, due to the "user generated content" (Krumm et al. 2008) property of social media and large volumes of noise. To some extent, this is likely to prolong the reaction time.

It is noticeable that the probabilities are consistently reduced from baseline models to weighted and weights-distributed models. Baseline models are established with an implicit assumption that only intentionally posted original contents (tweets) are considered important. In such cases, a higher level of correlation between the source and the receiver of information prompts more information being effectively transmitted (Das 2011). However, once the number of retweets is imposed as a weight on the models, there is a chance that a significant influence driven by key information is weakened due to the lack of retweets. Meanwhile, efficient market hypothesis requires a dense amount of non-redundant economic information (Giglio et al. 2008) that could only be presented in organic tweets, but could be weakened by a dense amount of retweets with insignificant contexts. Additionally, for a single tweet and its retweet, the reach and relevancy to the market is hard to measure and may not be the same. It could explain reduced sensitivities from weights distributed models, since the models take account of influences of unit information under an assumption that a tweet and its retweet are of the same quality. Moreover, the focus of this study is on Twitter, meaning that some information are not publicly available (or at least were not obtained on Twitter) with possible implications on minor market anomalies. Hence, lack of complete predictive sensitivity of unit information from an only source could happen at times. Furthermore, it is worth mentioning that a compromised linkage between keywords extraction model and LGBMs could occur and somewhat disturb predictions due to the nature of the model types. In such cases, the score of importance of a keyword (feature) would have a different meaning. Consequently, the feature with a higher feature importance in LGBM, that holds a high chance of purist split outcome and plays more significant role during predictions, might not be a better-scored keyword extracted from the document. Finally, the add-on weights of the number of retweets will make these differences more pronounced. The feature importance of each LGBM model is demonstrated in the Appendix.

Overall, our results suggest that when a classifier-depending tweet-sourced increase keywords make a prediction that an increase will happen during the next hour, 4 hours, or next day, the prediction is correct in about 78.06%, 84.63%, or 94.03% of all times, respectively. This indicates that daily information is more efficient than higher-frequency information regarding influences on the market. This result confirms previous findings that daily data are generally the most efficient (Vidal-Tomás 2022). It implies that with reduced percentages of incorrect predictions for a day, the investors taking a longer position based on the classifier's daily suggestions would be incurring less risk. However, the risk will increase on average 6.29% to 16.39% if investors take into account the same information from retweets, or even depend on a single random tweet or a retweet to rush their decision. In the worst case scenario, the correct prediction of increase for the next hour could dip around 60.42%, while most of the time the market price still shows efficient behaviors. Furthermore, the bearish market exhibits better incorporating changes with information, as probabilities based on original tweets hover around 83.08%, 87.77%, and 94.60% correctness for predicting subsequent hourly, 4 hourly, and daily decreases, respectively. Negative information revealed in retweets do not either play the same role as tweets for signaling price changes. However, market changes do show slightly different patterns while reflecting daily based information. Specifically, the bullish market incorporates daily information available from both unit and whole tweets and their retweets better than bearish market by about 0.89% ∼ 3%, as in other daily model bearish market takes over by a negligible 0.57%. These observations confirm the importance of daily information to the EMH, implying that negative and positive movements could be equally accurately predicted by public information available in daily Bitcoin market. Reflection to the whole volume of positive information is even stronger. These findings are in line with previous work on market efficiency (Bartos et al. 2015; Nadarajah and Chu 2017; Tiwari et al. 2018) and particularly in Bitcoin market (Kraaijeveld and De Smedt 2020; Pant et al. 2018; Shen et al. 2019; Choi 2021).

## 5. Model Comparisons and Robustness Tests

We include two types of robustness tests for confirming consistencies of results. Gradient Boosting Decision Tree (GBDT) models without GOSS and EFB are deployed to test whether results remain homogeneous as two advanced techniques, boosting type and bundling based algorithm, are replaced in decision tree models. In addition, to eliminate biases from different proportional splits, we test the same proposed LGBM models on

Table 6: Results of Probability and Adjusted Probability of Market Movements Prediction by Robustness Tests with GBDT

| Interval Feature Set | Baseline Model | | | | Weighted Model | | | | Weights Distributed Model | | | |
| --- | --- | --- | --- | --- | --- | --- | --- | --- | --- | --- | --- | --- |
| | $P_{Increase}$ | | $P_{Decrease}$ | | $P_{Increase}$ | | $P_{Decrease}$ | | $P_{Increase}$ | | $P_{Decrease}$ | |
| | *Prob* | *AdjP* | *Prob* | *AdjP* | *Prob* | *AdjP* | *Prob* | *AdjP* | *Prob* | *AdjP* | *Prob* | *AdjP* |
| *Hk**In* | 79.07% | 77.09% | 21.74% | 21.20% | 73.09% | 71.26% | 28.85% | 28.13% | 61.69% | 60.15% | 39.93% | 38.94% |
| *Hk**De* | 15.77% | 15.38% | 84.20% | 82.10% | 25.24% | 24.61% | 74.15% | 72.30% | 34.38% | 33.52% | 64.25% | 62.65% |
| *4Hk**In* | 84.28% | 83.87% | 15.81% | 15.74% | 75.94% | 75.57% | 25.02% | 24.90% | 67.15% | 66.83% | 34.57% | 34.41% |
| *4Hk**De* | 10.74% | 10.69% | 87.72% | 87.30% | 22.71% | 22.60% | 76.42% | 76.05% | 31.16% | 31.01% | 69.48% | 69.14% |
| *Dk**In* | 95.87% | 95.53% | 3.89% | 3.87% | 90.38% | 90.07% | 14.15% | 14.10% | 81.46% | 81.17% | 18.49% | 18.43% |
| *Dk**De* | 4.25% | 4.23% | 96.01% | 95.67% | 16.00% | 15.94% | 86.24% | 85.94% | 15.66% | 15.61% | 81.88% | 81.60% |

Table 7: Results of Probability and Adjusted Probability of Market Movements Prediction By Robustness Tests with 70-30 Splits

| Interval Feature Set | Baseline Model | | | | Weighted Model | | | | Weights Distributed Model | | | |
| --- | --- | --- | --- | --- | --- | --- | --- | --- | --- | --- | --- | --- |
| | $P_{Increase}$ | | $P_{Decrease}$ | | $P_{Increase}$ | | $P_{Decrease}$ | | $P_{Increase}$ | | $P_{Decrease}$ | |
| | *Prob* | *AdjP* | *Prob* | *AdjP* | *Prob* | *AdjP* | *Prob* | *AdjP* | *Prob* | *AdjP* | *Prob* | *AdjP* |
| *Hk**In* | 79.93% | 77.93% | 21.14% | 20.61% | 74.07% | 72.22% | 27.82% | 27.13% | 61.95% | 60.40% | 39.65% | 38.66% |
| *Hk**De* | 14.91% | 14.54% | 85.07% | 82.94% | 25.01% | 24.38% | 74.76% | 72.90% | 34.15% | 33.30% | 65.36% | 63.73% |
| *4Hk**In* | 85.09% | 84.68% | 15.34% | 15.26% | 75.62% | 75.25% | 24.63% | 24.52% | 67.36% | 67.03% | 33.57% | 33.41% |
| *4Hk**De* | 10.73% | 10.68% | 88.71% | 88.28% | 22.94% | 22.83% | 77.62% | 77.24% | 30.65% | 30.50% | 70.12% | 69.78% |
| *Dk**In* | 94.65% | 94.32% | 6.90% | 6.87% | 86.86% | 86.55% | 18.72% | 18.65% | 80.79% | 80.51% | 22.60% | 22.52% |
| *Dk**De* | 6.22% | 6.19% | 94.80% | 94.47% | 19.67% | 19.60% | 84.61% | 84.31% | 18.99% | 18.92% | 81.25% | 80.96% |

$P_{Increase}$, $P_{Decrease}$, *Prob*, *AdjP*, **Interval Feature Set**, *In* and *De* hold the same explanations as in the Table 5.

training and testing set with 70-30 splits and also expect similar results obtained by different splits. Results of two types of robustness tests are reported in Table 6 and Table 7, respectively.

Results in both robustness tests yield the same pattern as in the proposed model. Namely, regardless of different models, predictions towards daily price movements hold the highest probability. Also, probabilities are consistently high in the baseline model despite variant time intervals. In general markets have stronger reactions towards negative information than positive information. The only exceptional case happens in the weighted daily models that also persistently occurs in our proposed LGBM studies. These findings confirm that the predictions are robust and consistent in spite of the tuning of hyper-parameters and ratios of validation splits. Furthermore, we compare results from proposed LGBMs with results from robustness tests and present them in Table 8.

Among three groups of tests, in 39% cases, probabilities predicted in proposed LGBMs are higher than others. The difference is within 3.5% between the highest number obtained by other models and proposed LGBMs. It is not unexpected that robustness test boosted by GBDT performs better than LGBMs with GOSS booster and EFB in the daily interval cases. Since the size of dataset in daily cases is significantly smaller than others, the advantage of GOSS and EFB from reduced computation of smaller gradients and whole sets of features could not benefit training procedure. Oppositely, GBDT taking account of the contributions of all gradients towards information gain and features may have higher accuracy with smaller datasets. In additional, differences between 80-20 splits and 70-30 splits in LGBMs with the same setup of hyper-parameters are negligible (below 0.5%), proving the validity of proposed LGBMs. To keep the study consistence, we conclude final results according to proposed LGBMs with 80-20 splits. We also suggest that investing decisions about daily market could also refer to results from GBDT.

Table 8: Results of Adjusted Probability of Market Movements Prediction By Proposed LGBMs and Robustness Tests

| | ProLGBMs | Robustness Test | | ProLGBMs | Robustness Test | | ProLGBMs | Robustness Test | |
|---|---|---|---|---|---|---|---|---|---|
| **Hyper-Para** | *GOSS, +EFB* | *GBDT, -EFB* | *GOSS, +EFB* | *GOSS, +EFB* | *GBDT, -EFB* | *GOSS, +EFB* | *GOSS, +EFB* | *GBDT, -EFB* | *GOSS, +EFB* |
| **Split** | *80-20* | *80-20* | *70-30* | *80-20* | *80-20* | *70-30* | *80-20* | *80-20* | *70-30* |
| **IFSet** | Baseline Model | | | Weighted Model | | | Weights Distributed Model | | |
| *Hk**In*** | 78.06% | 77.09% | 77.93% | 72.26% | 71.26% | 72.22% | 60.42% | 60.15% | 60.40% |
| *Hk**De*** | 83.08% | 82.10% | 82.94% | 73.03% | 72.30% | 72.90% | 63.46% | 62.65% | *63.73%* |
| *4Hk**In*** | 84.63% | 83.87% | *84.68%* | 75.99% | 75.57% | 75.25% | 66.97% | 66.83% | *67.03%* |
| *4Hk**De*** | 87.77% | 87.30% | *88.28%* | 76.99% | 76.05% | *77.24%* | 70.32% | 69.14% | 69.78% |
| *Dk**In*** | 94.03% | *95.53%* | 94.32% | 86.50% | *90.07%* | 86.55% | 81.06% | *81.17%* | 80.51% |
| *Dk**De*** | 94.60% | *95.67%* | 94.47% | 83.50% | *85.94%* | 84.31% | 80.17% | *81.60%* | 80.96% |

\* **ProLGBMs**: proposed LGBMs; **Hyper-para**: hyper-parameters; **+(-)EFB**: with (without) EFB
\* **IFSet** (Interval Feature Set); ***In*** and ***De*** hold same explanations as in previous tables.
\* All numbers are the adjusted probability acquired by proposed LGBMs and robustness tests, responding to expected increasing/decreasing price in next time interval.
\* Highest probabilities predicted in robustness tests instead of proposed LGBMs are formatted in italic.

## 6. Conclusion

In the present study, we suggest a framework to measure and quantify the validity of the EMH in Bitcoin market when different market reaction times are considered. We selected a complete sample of 28,739,514 qualified tweets related to the topic "Bitcoin" spanning a 5-year interval and implemented distance measurements of semantic vector spaces calculated by GloVe model to address the fundamental dissimilarities of information prior to increasing and decreasing market movements. Integrating YAKE keywords extraction model and feature encoding mechanisms, we encoded data with 64 keywords and their ranked scores extracted in an hour, 4 hours, and a day, separately before the largest increasing and decreasing movements. Following that, we developed three groups of separate classifiers with LGBMs, including taking into account the reach of diffused information and the influence of unit information, and identified the efficiency of market reactions corresponding to different scenarios. After 10-fold cross-validation on an 80-20 split, we tested each final model on 20% of randomly selected subsamples with binary labels, increase or decrease, to generate a basic perspective on how accurately each model can distinguish different market movements and to finalize the analysis of probabilities of each set of increase/decrease keywords being correctly predicted as reflected in market movements. The results are adjusted by the baseline distance weights and show that in the best scenarios, classifier predictions demonstrate 78.06% (83.08%), 84.63% (87.77%), and 94.03% (94.60%) for the following increase (decrease) in one hour, 4 hours, and one day, suggesting that prices are adjusted to reflect publicly available and organic information of 78.06% (83.08%), 84.63% (87.77%), and 94.03% (94.60%) of certain time intervals in bull (bear) markets, respectively. We compared differences of probabilities predicted by diffused information and unit information across different time intervals and explained that they could be caused by the causality of delayed market reactions, the reach of retweets, the unavailable information, and market anomalies.

Our research confirms the presence of semi-efficient market efficiency in Bitcoin and also indicates that information from organic tweets is the most reliable in predicting upcoming market movements. Investors would benefit the most from information one day prior to taking an action in the market and accrue more risk if their investment decision would depend on information 1 hour or 4 hours in advance. The information revealed from retweets does not carry the same levels of reliability as the same information obtained from tweets with respect to predictions of subsequent price changes. Additionally, potential investors should take into account the overall popularity of an informative keyword instead of focusing on early signals released from a random tweet/retweet for decision making. Finally, our study was applied to the entire market without differentiating the trades initiated by a specific trader group, such as professional institutions or individual traders. Information from various sources (traders) could have different effects and time lags towards market reactions. Therefore, to avoid disturbances from informative biases, it is advisable to take into consideration the movements of the entire market instead of a price spike or a dip in very short amount of time caused by increased trading volumes from a single trader group.

Our work can be extended in different directions. First, we have found that the validity of the EMH improves as testing time windows are extended in both bull and bear markets. However, it would be interesting to know different reasons that cause inefficient reactions to appear in 1-hour or 4-hour time windows. It is also worthy to explore possible explanations why the EMH cannot persist for the remaining $\sim 5\%$ of days. We assume that some heterochthonous events could be the source of biases in the prediction models. Thus, an event study could be beneficial for the future research. In this study, due to the limits on availability of trusted sources of historical market data at minute resolution, we constrained our analysis to 1-hour resolution. However, shorter time intervals, such as reactions in 1 minutes, 5 minutes, etc., should be studied in the future in order to capture the market reactions to significant events such as unexpected announcements in the media or shock at the price movements. In these cases, we could expect more accurate results and stronger efficient market reactions on shorter timescales than one day.

We also notice that some metrics of tweets could be tested for purposes of improving performances of weighted and weights-distributed models. For instance, the number of "like" and "quoted" of a tweet may play a role on deciding the popularity of important keywords. However, since we did not have a clear metric according to which we could distinguish which attributes of each Twitter could generate most significant influences in mass-communication, it is important to mention that considering only the number of "retweet" might be not sufficient for making investing decisions. A further complication are potential interactions between different Twitter attributes. Thus, the interactions of attributes could be considered in the future in order to improve prevalence and accuracy of the model. A particular topic possibly warranting further study is the impact of a potential snowball effect of large number of followers, retweets, and other attributes possessed by the same tweet, on predicting the influence of organic tweets. In addition, the study could be extended with a different number of keywords that may yield additional fruitful outcomes. It is worth mentioning that different sources of market data could be applied in future studies and may provide more convincing results. Moreover, it would be informative and timely to extend our study to other cryptocurrencies whose market behaviors and reactions to the public information (Twitter) might be very different. Thus, the EMH within the scope of each significant modern cryptocurrency should be studied individually before a more general conclusion regarding the market actions of cryptocurrencies as a new class of asset could be made.

# Appendix A. Feature Importance Reports of LGBM Models

Figure A.5: Feature Importance in Baseline LGBM models: Increase (Left) and Decrease (Right) Keywords within Hourly Interval

Figure A.6: Feature Importance in Baseline LGBM models: Increase (Left) and Decrease (Right) Keywords within the Interval of 4 Hours

Figure A.7: Feature Importance in Baseline LGBM models: Increase (Left) and Decrease (Right) Keywords within Daily Interval

Figure A.8: Feature Importance in Weighted LGBM models: Increase (Left) and Decrease (Right) Keywords within Hourly Interval

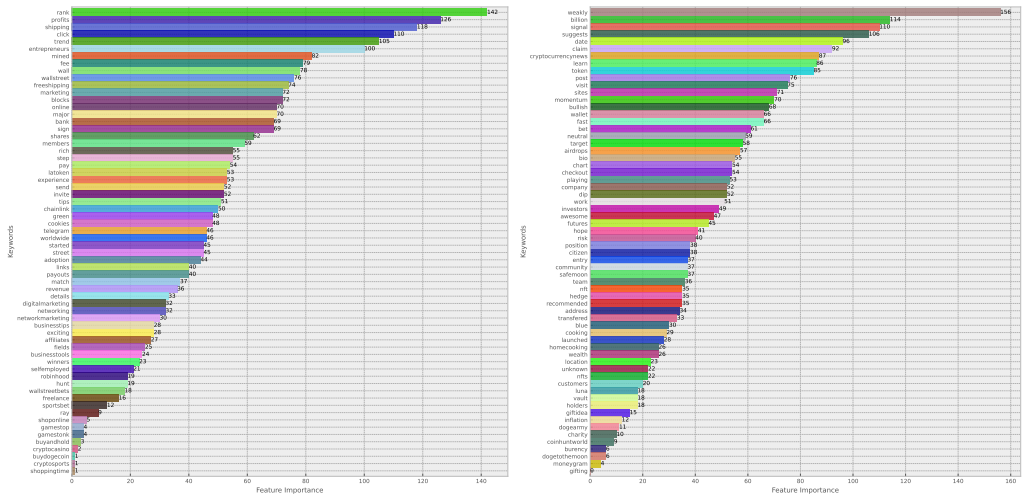

Figure A.9: Feature Importance in Weighted LGBM models: Increase (Left) and Decrease (Right) Keywords within the Interval of 4 Hours

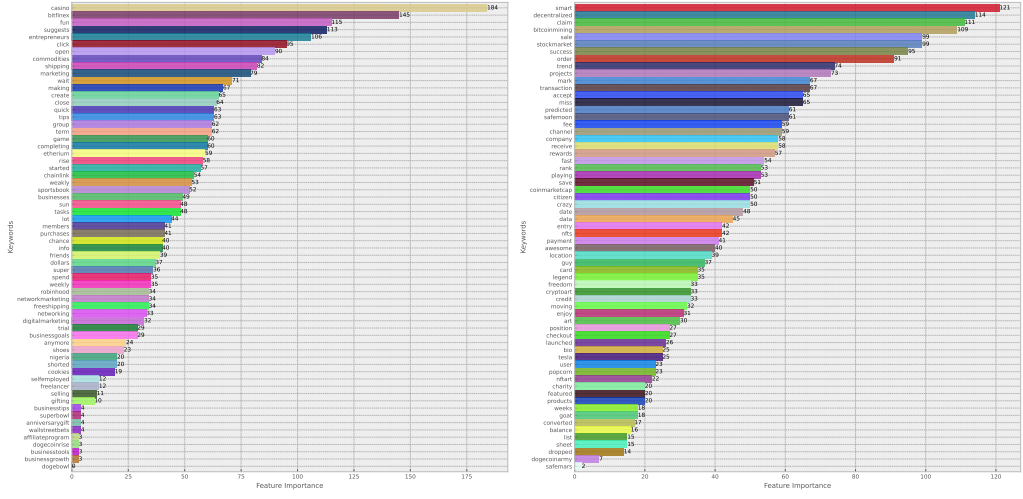

Figure A.10: Feature Importance in Weighted LGBM models: Increase (Left) and Decrease (Right) Keywords within Daily Interval

Figure A.11: Feature Importance in Weights Distributed LGBM models: Increase (Left) and Decrease (Right) Keywords within Hourly Interval

Figure A.12: Feature Importance in Weights Distributed LGBM models: Increase (Left) and Decrease (Right) Keywords within the Interval of 4 Hours

Figure A.13: Feature Importance in Weights Distributed LGBM models: Increase (Left) and Decrease (Right) Keywords within Daily Interval